\documentclass[pageno]{jpaper}

\usepackage[normalem]{ulem}
\usepackage{graphicx}
\usepackage{tabularx}
\usepackage{adjustbox}
\usepackage{listings}
\usepackage{float}
\usepackage{parcolumns}
\usepackage{pdfpages}
\usepackage{xcolor}
\usepackage{xurl}
\usepackage{todonotes}
\usepackage{authblk}
\RequirePackage{glossaries}
\RequirePackage{tcolorbox}

\graphicspath{ {./figures/} }
\DeclareGraphicsExtensions{.png,.pdf}

\newacronym{MMU}{MMU}{Memory Managment Unit}
\newacronym{TLB}{TLB}{Translation Lookaside Buffer}
\newacronym{MPU}{MPU}{Memory Protection Unit}
\newacronym{PMP}{PMP}{Physical Memory Protection}
\newacronym{LUT}{LUT}{Look-Up Table}
\newacronym{IPC}{IPC}{Inter Process Communication}
\newacronym{OS}{OS}{Operating System}
\newacronym{EOS}{EOS}{Embedded Operating System}
\newacronym{RTOS}{RTOS}{Real Time Operating System}
\newacronym{DoS}{DoS}{Denial of Service}
\newacronym{CHERI}{CHERI}{Capability Hardware Enhanced RISC Instructions}
\newacronym{AWS}{AWS}{Amazon Web Services}
\newacronym{LLVM}{LLVM}{Low Level Virtual Machine}
\newacronym{FPGA}{FPGA}{Field Programmable Gate Arrays}
\newacronym{TCP}{TCP}{Transmission Control Protocol}
\newacronym{IP}{IP}{Internet Protocol}
\newacronym{IoT}{IoT}{Internet of Things}
\newacronym{API}{API}{Application Programming Interface}
\newacronym{OTA}{OTA}{Over The Air}
\newacronym{ACL}{ACL}{Access Control List}
\newacronym{TCB}{TCB}{Trusted Computing Base}
\newacronym{TLS}{TLS}{Thread Local Storage}

\newcommand{\note}[3]{{\color{#3}[ #1---#2 ]}}
\renewcommand{\note}[3]{}

\begin{document}

\title{CompartOS: CHERI Compartmentalization for Embedded Systems}
\author[1]{Hesham Almatary}
\author[1]{Michael Dodson}
\author[1]{Jessica Clarke}
\author[1]{Peter Rugg}
\author[1]{Ivan Gomes}
\author[2]{Michal Podhradsky}
\author[3]{Peter G. Neumann}
\author[1]{Simon W. Moore}
\author[1]{Robert N. M. Watson}
\affil[1]{University of Cambridge}
\affil[2]{Galois, Inc.}
\affil[3]{SRI International}


\date{}
\maketitle

\thispagestyle{empty}


\begin{abstract}
Existing high-end embedded systems face frequent security attacks. Software compartmentalization is one technique to limit the attacks' effects to the compromised compartment and not the entire system.
Unfortunately, the existing state-of-the-art embedded hardware-software solutions do not work well to enforce software compartmentalization for high-end embedded systems. MPUs are not fine-grained and suffer from significant scalability limitations as they can only protect a small and fixed number of memory regions. On the other hand, MMUs suffer from non-determinism and coarse-grained protection.

This paper introduces CompartOS as a lightweight linkage-based compartmentalization model for high-end, complex, mainstream embedded systems. CompartOS builds on CHERI, a capability-based hardware architecture, to meet scalability, availability, compatibility, and fine-grained security goals.

Microbenchmarks show that CompartOS' protection-domain crossing is 95\% faster than MPU-based IPC.
We applied the CompartOS model, with low effort, to complex existing systems, including TCP servers and a safety-critical automotive demo.
CompartOS not only catches 10 out of 13 FreeRTOS-TCP published
vulnerabilities that MPU-based protection (e.g., uVisor) cannot catch but can also recover
from them. Further, our TCP throughput evaluations show
that our CompartOS prototype is 52\% faster than relevant MPU-based compartmentalization models (e.g., ACES),
with a 15\% overhead compared to an unprotected system.
This comes at an FPGA's LUTs overhead of 10.4\% to support CHERI for an unprotected baseline RISC-V processor,
compared to 7.6\% to support MPU, while CHERI only incurs 1.3\% of the registers area overhead compared to 2\% for MPU.
\end{abstract}

\section{Introduction and Related Work}


The inexorable need to add more features and leverage connectivity in embedded systems
creates potential attack vectors~\cite{Costin2014, cui2010quantitative,
    Nicoul2011, AnandaKumar2014, Virat2018, koscher2010experimental, Checkoway2011, Miller2014}
in areas that were not subject to security concerns before.
Individual's privacy could now be violated, cars and planes could crash, credit-card details
could be stolen, and  medical devices could critically malfunction, affecting vital life-concerning
actions or leak sensitive patients' details. The prevalence of such devices creates large scale risks
for the economy, national security, and the safety of large populations.

Embedded systems and \gls{IoT} get more complex and feature-rich every day. While complexity
increases, they tend to maintain the real-time requirements and determinism of previous generations.
We are narrowing down the scope of systems we are targeting,
to the mainstream, large, feature-rich systems that are unprotected and require some form of security
to defend against unknown and future vulnerability-based software attacks.
That is, we do not try to provide a security model or solution for small embedded systems
that get created from scratch.
To protect such systems from the inevitable vulnerabilities exploitable in the software, all
protection models require some hardware support, such as privilege rings, \gls{MPU}s, TrustZone~\cite{enwiki:1069468401}, or \gls{MMU}s.
However, there is currently
a gap in hardware-software security for targeted embedded applications; the M-class-like processors~\cite{enwiki:1070148237}
with \gls{MPU} and TrustZone
are too small and inadequate for their needs in terms of scalability, memory requirements, and fine-grained security,
while A-class-like processors~\cite{enwiki:1061921480} and \gls{MMU}s do not meet the fine-grained security or real-time and determinism requirements. Thus,
such systems tend to stay unprotected or apply complicated workarounds and manual task-based compartmentalization to get some form of security
with non-trivial development, maintenance, and time-to-market overheads.
The scope such systems fall in is safety-critical avionics and automotive.

Based on the security analysis of embedded systems in~\cite{Papp2015, parameswaran2008embedded, kocher2004security},
vulnerability-based software attacks (e.g., buffer overflows) are the most common and
they are what we are concerned about in this paper.
\textit{Software compartmentalization}~\cite{watson2015cheri, gudka2015clean, karger1987limiting,
    provos2003preventing, kilpatrick2003privman, watson2010capsicum} is a technique to split up a large
monolithic software into smaller compartments in order to reduce the attack vector and limit the
effects of a successful software attack only to the compromised compartment.
There have been multiple recent attempts~\cite{sensaoui2019depth} to provide software compartmentalization,
sandboxing, and isolation in
embedded systems by relying on \gls{MPU}s as a standard hardware feature for embedded
systems. Due to the hardware limitations of \gls{MPU}s (by design)~\cite{zhou2019good},
such solutions are not scalable and do not provide fine-grained protection, especially
for rich and complex mainstream software stacks with hundreds of thousands of lines of code (LoC) and many resources.
For example, the state-of-the-art deployed \gls{RTOS}es such as FreeRTOS-MPU~\cite{freertosmpu21}, TockOS~\cite{levy2017multiprogramming}, and
Mbed uVisor~\cite{uvisorgithub22} rely on an MPU-based task or process model for isolation. This could work fine for relatively small
embedded systems with a small number of resources, but not for large systems
as \gls{MPU} hardware scales poorly with tens or hundreds of
compartments and resources that need to be isolated while maintaining some form of determinism (hence cannot use MMUs either).
Further, for mainstream applications to be secured
using such technologies, they will have to be ported to use new \gls{API}s or memory-safe
languages (e.g., Rust) and redesigned around tasks or processes. This could be
a deal-breaker for large and mainstream projects that are written in different languages and
not designed around tasks or processes due to the time-to-market requirements and development
overheads.
ACES~\cite{clements2018aces} tries to provide an automatic compartmentalization technique that is
source-file or IO-based for small embedded systems, using the \gls{MPU}. This helps with the compatibility
requirement by not requiring to port or refactor mainstream code. While promising for small and simple embedded
systems, ACES still cannot target large software projects due to the \gls{MPU} hardware limitations.
Further, ACES does not try to provide an \gls{OS} protection model, including threading, secure interrupt
handling, or dynamic memory allocation.
TrustLite~\cite{Koeberl2014} and TyTan~\cite{Brasser2015} try to mitigate the \gls{MPU} design limitations
by increasing the number of memory regions to protect. However, increasing the \gls{MPU} regions does not scale
in either hardware due to the increased associative lookups or in software as context switches will
incur O(N) performance overhead where N is the number of memory regions to protect, per protection domain.
In almost all of the MPU-based protection published literature, the limitations of \gls{MPU} scalability
are admitted, and, therefore, they only try to target small systems. Further, none of these works attempts
to address availability, an essential requirement of safety-critical systems.
This leaves large, mainstream and complex embedded systems that require some form of security
while maintaining determinism as a real-time requirement still vulnerable.\\

CompartOS aims to fill this security requirement gap in mainstream, complex, high-end embedded systems,
where manually porting them to use \gls{MPU} or \gls{MMU} hardware, \gls{OS}' \gls{API}s, and/or rewriting them in memory-safe languages is
inadequate or requires significant redesign and reimplementation efforts which would have a high development,
testing, and deployment cost and are error-prone.
More specifically, CompartOS aims to reduce the effects of programming errors and vulnerabilities
between software components running on the same embedded processor, in the same address space and privilege ring
and sharing resources by applying a novel software compartmentalization technique that builds on \gls{CHERI}. Compatibility,
availability, and scalability are the main design goals in this work.

The primary research contribution of this paper is the CompartOS model (and its evaluation via prototyping)
as a linkage-based, automatic compartmentalization model for embedded systems.
Building on \gls{CHERI} to provide software compartmentalization protects against memory-safety attacks that
represent 70\% of software vulnerabilities in commodity \gls{OS}es according to Microsoft~\cite{cimpanu_2019}.
However, in embedded systems, it is not enough to catch a security violation but to appropriately handle it
to maintain the integrity and availability of security requirements.
In terms of contributions, CompartOS, as a model for CHERI-enabled, \gls{MMU}-less, embedded systems,
is comparable to the well-defined UNIX process model
(inspired by Multics~\cite{saltzer1974protection, daley1968virtual})
for general-purpose systems.

CompartOS is novel and differs from state-of-the-art systems. First, CompartOS significantly differs from previous \gls{CHERI} work as follows:
\begin{itemize}
    \item Unlike~\cite{watson2015cheri, Davis2019}, CompartOS focuses on embedded systems by enforcing complete protection and compartmentalization solely using \gls{CHERI} (i.e., by replacing \gls{MMU}s and \gls{MPU}s) instead of
    being complementary to the \gls{MMU}-based UNIX process model. Further, CompartOS leverages system-level compartmentalization, rather than just application-level compartmentalization, on all OS aspects, including kernels,
    device drivers, OS libraries, and applications.
    \item Unlike CheriOS~\cite{esswood2021cherios} and CheriRTOS~\cite{xiaCheriRTOSCapabilityModel2018} (where both are task-based), CompartOS does not introduce a new OS or \gls{API} to manually compartmentalize complex software, but it automatically enforces compartmentalization based
    on the linkage model and programming languages.
    \item Focuses on availability and recovery in safety-critically embedded systems after catching \gls{CHERI} faults (or others),
    rather than accepting fail-stop on, for example, a memory protection fault.
\end{itemize}

Second, unlike MPU-based state-of-the-art secure embedded architectures~\cite{uvisorgithub22, Kim2018, clements2018aces, levy2017multiprogramming, Koeberl2014, Brasser2015, sensaoui2019depth}, CompartOS:
\begin{itemize}
    \item Scales to hundreds of compartments and resources without hardware protection limitations.
    \item Targets existing mainstream and complex unprotected embedded (operating) systems and libraries that could scale to millions of lines of code.
    \item Provides capability-based security that is enforced on every pointer and linkage module, inter- and intra- task/application.
    \item Assumes malicious compartments and aims to protect against current and unknown future software vulnerabilities.
    \item Outperforms comparable state-of-the-art systems.
\end{itemize}

In summary, we describe the major contributions in this paper as follows:

\begin{description}
    \item[CompartOS abstract model] A generic CHERI-based compartmentalization and protection model for embedded (operating) systems, which can be effortlessly applied
    to most mainstream embedded systems and \gls{RTOS}es.

    \item[CheriFreeRTOS implementation of CompartOS] A real-world, sound prototype of the CompartOS model in FreeRTOS including a secure loader and
    protecting the kernel, OS/IoT libraries, and large and complex deployed mainstream applications.

    \item[MPU-based comparisons] Implementations of the state-of-the-art MPU-based protection models in FreeRTOS to fairly evaluate the CheriFreeRTOS prototype against, at scale.

    \item[Evaluation] Demonstrating complete and functional real-world, deployed use cases with CompartOS and other MPU-based models and providing security, compatibility, availability, and performance analysis.
\end{description}

\section{Background}

This section provides the
background necessary to contextualize subsequent use of CHERI,
and the run-time linker (\texttt{libdl}) in CompartOS.


\subsection{CHERI}

CHERI is an architecture-neutral, state-of-art hardware capability system that
provides the architectural features necessary to support efficient and
fine-grained memory protection and software-defined
compartmentalization~\cite{watson2015cheri}.
CHERI-aware Instruction Set Architecture (ISA) extensions and deployments exist
for MIPS, RISC-V, and Armv8
(Morello)~\cite{armlimitedArmArchitectureReference2020} hardware and software
stacks.  CHERI is designed to enforce the principles of least privilege~\cite{Saltzer1975} and
intentional use.  The former is enforced by explicitly
granting to a software component only the privileges required to perform its
intended service.  The latter is enforced by requiring
that memory operations, capability manipulations, and privileged operations
explicitly provide and name capabilities holding restricted permissions to
perform a service. This avoids the ambient authority that can produce a
confused deputy~\cite{Boebert1996}.
There are two CHERI protection models actively being researched:
pointer safety and software compartmentalization.

Pointer safety in C/C++ languages has been the
primary application of CHERI to date. It leverages the
compiler toolchain to map C/C++ pointers to CHERI's capabilities. Every pointer is
represented as a CHERI capability with base, bounds, and permissions, and CHERI-aware
instructions enforce complete spatial pointer safety.
Pointers include conventional data pointers, arrays, function pointers, stack and heap allocations, and
return addresses. Violations trap to the executive, which determines how to handle the
memory fault.

Software compartmentalization is a newer application of CHERI.
The purpose of compartmentalization is to divide a monolithic system into smaller components, where
each component is granted only the explicit privileges necessary to perform its service.
Compartmentalization reduces the system's attack surface, limits the effect of a successful exploit
to the affected compartment, protects against currently-unknown vulnerabilities in a software
component, and provides opportunities to improve fault handling due to the well-defined boundaries
between compartments.
The main contribution of this paper is a novel CHERI-based compartmentalization model for embedded systems.

\subsection{LIBDL}

\label{libdl}

\texttt{libdl}~\cite{libdlrtems} is a component
initially developed by the RTEMS project~\cite{rtems} to dynamically load and link ELF modules
against a single-address-space, MMU-less, statically-linked ELF image. After the loading/linking
process is completed, the resulting performance is the same as if the ELF image was generated at
build-time with static linkers. \texttt{libdl} is useful in patching, fixing bugs, updating
libraries, and adding more features (e.g., applications, drivers, and libraries) without having to
reset the entire system, affecting its availability, as in conventional software updates.

\section{Threat Model}
Compartments are untrusted and could attempt to violate (deliberately or unintentionally)
language-based memory-safety guarantees enforced by CHERI, OS-level isolation mechanisms, or
externally defined inter-compartment security policies. Like other isolation technologies, we
assume the attacker seeks to compromise the system's confidentiality and integrity.  Unlike other
technologies, we also assume the attacker seeks to disrupt system availability, and we explicitly
design and evaluate against this target.
We assume a CHERI-aware attacker that is either resident on the device (e.g., controlling a library) or in direct communication with a vulnerable software comopnent on the device (e.g., the TCP/IP library) but has not compromised the TCB.


\section{Design Principles and Requirements}

\label{sec:requirements}

This section discusses the requirements and goals that CompartOS is designed to meet and how they are
useful, and how they compare to other state-of-the-art embedded security architectures. It also briefly
describes how CompartOS meets those requirements, but the thorough evaluations are further discussed
in subsequent sections.

\paragraph{Capability-Based Security:}
\label{design:cap_security}
Due to their OS design and limited hardware, embedded systems generally lack
an access control model such as \gls{ACL} and filesystems in UNIX-based
\gls{OS}es. This means that nothing stops an attacker on an embedded system
from accessing the entire memory space and all system resources.
Capability-based access control provides a scalable, lightweight mechanism
that can be implemented with little perturbation to existing software.
CompartOS aims to leverage capability-based security~\cite{miller2003capability,
    levy2014capability, shapiro1999eros} everywhere: not only protecting kernel objects (as in
capability-based microkernels~\cite{klein2009sel4}) but also providing hardware memory protection
and programming-language pointer safety. That is, CompartOS extends security enforcement into the
compartment itself, rather than just providing isolation guarantees between compartments.


\paragraph{Scalable Automatic Compartmentalization:}
\label{design:scalablity}
Composing multiple components, libraries, or applications
is increasingly common in embedded systems, with third-party libraries and standards
becoming more feature-rich every day.
Therefore, increasing the number of resources
and features (represented as compartments) should
be scalable, without significantly affecting performance and security guarantees or
requiring major engineering and maintenance overhead.

\paragraph{Source-Code Compatibility:}
\label{design:compatability}
Mainstream, large, complex, and unprotected embedded system software could be written
in millions of lines of code. Providing security for such systems at a minimum
development and maintenance overhead is one of the main requirements.
That is, applications should require few, if any, source-code changes.
We deliberately chose to avoid designing a completely new research OS or
rewriting existing non-secure systems in memory-safe languages such as Rust,
all of which complicate learning,
development, runtime, and maintenance overheads, and can be a deal-breaker for
large complex mainstream systems in the industry.

\paragraph{Improved Availability:}
\label{design:availablity}
Many attacks target the availability of embedded systems rather than
confidentiality and integrity.
General-purpose OSes (e.g., UNIX) have a lower focus on availability
and it is often a secondary design goal.
In contrast, breaking availability in safety-critical embedded systems could sometimes cost lives,
e.g., if they manage medical databases, information on air traffic, and so on.


\paragraph{Real-time and Embedded-Systems Requirements:}
\label{design:realtimeness}
Most mainstream \gls{RTOS}es still need some form of determinism
as a real-time requirement, and that is why they avoid using an \gls{MMU}
to enforce security and protection.
CompartOS does not introduce non-determinism or add to the complexity of the
to-be-compartmentalized real-time embedded system itself.

\section{Compartmentalization Model}
\label{compartos_model}

\begin{figure}[]
    \centering
    \includegraphics[width=\columnwidth,height=4cm]{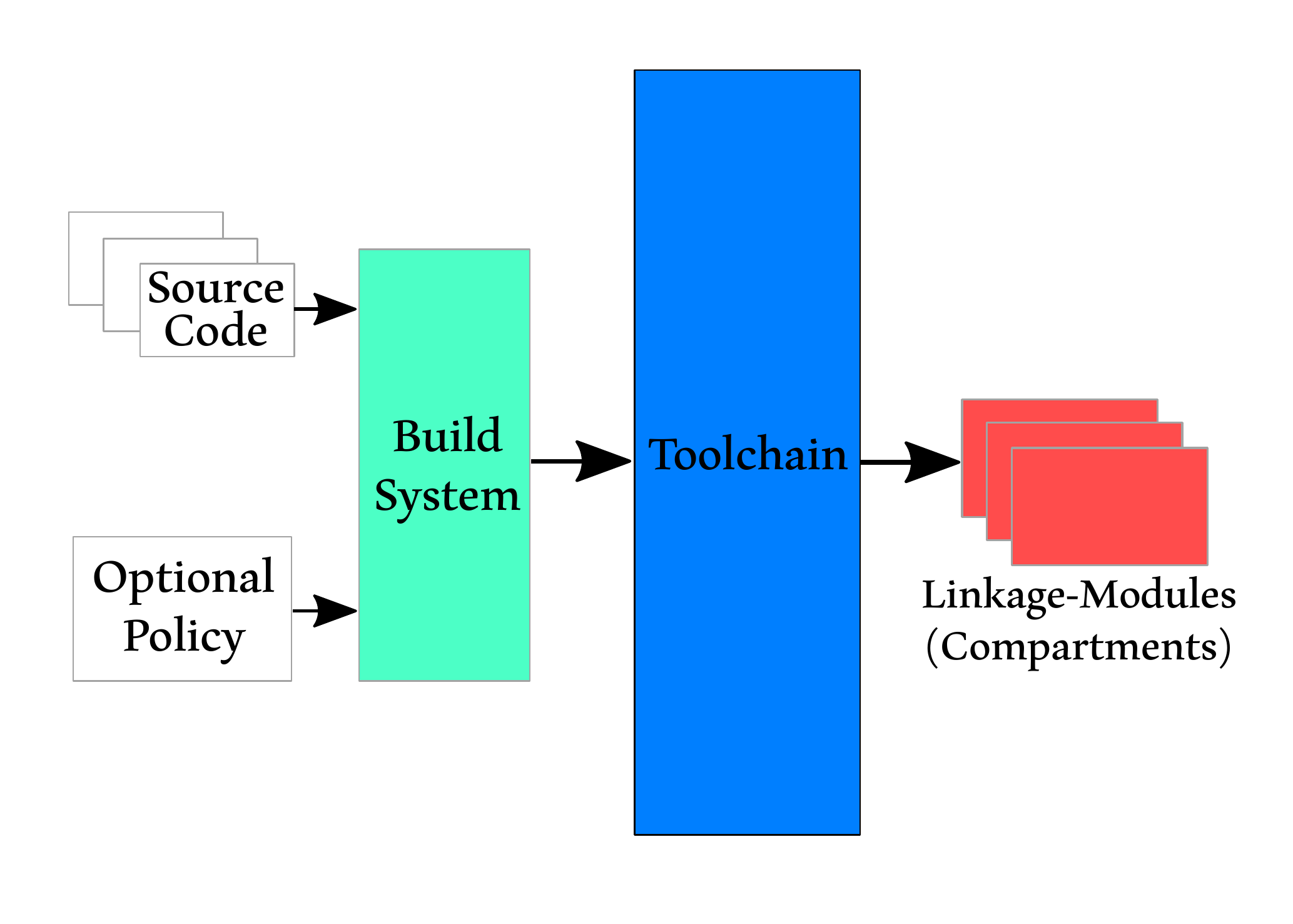}
    \caption{CompartOS build-time workflow.
        Developers feed the (mainstream) source-code
        to the build system as well as an optional security policy that splits up the source-code
        into logical compartments with well-defined \gls{API}s (e.g., libraries), along with the restricted communications between them.
        The output is linkage modules that are fed to the dynamic secure loader at runtime.}
    \label{fig:comprtos_tooling}
\end{figure}

\begin{figure}[]
    \centering
    \includegraphics[width=\columnwidth,height=7cm]{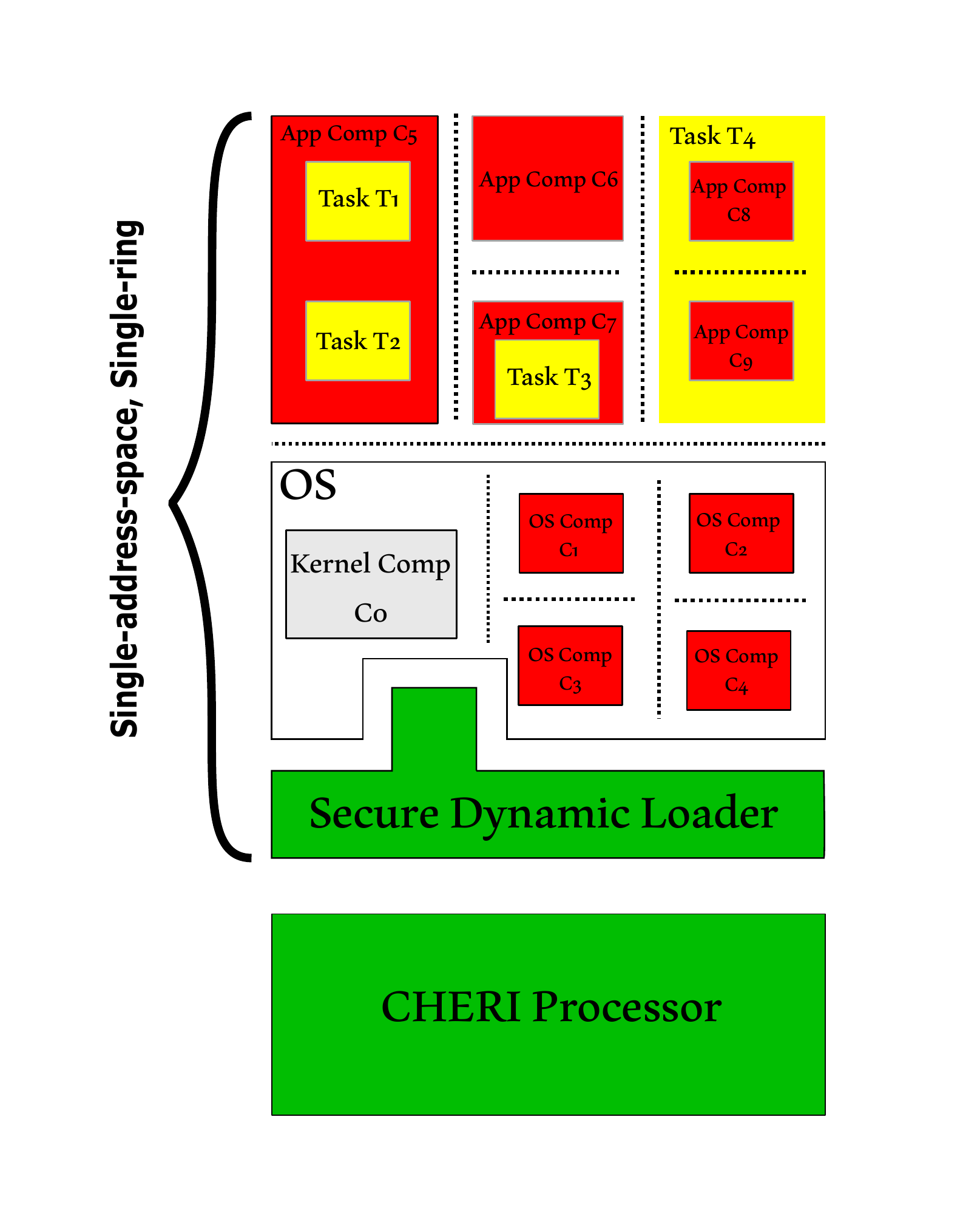}
    \caption{CompartOS runtime model. Green boxes are trusted. Everything else is untrusted, including the toolchain. Linkage module compartments
         are in red, and OS tasks (or threads) are in yellow.
        Dotted lines indicate isolation boundaries
        with potential bridges for communication between compartments as specified by the security policy.
        The secure loader notch indicates some form of integration with the OS (e.g., threading).}
    \label{fig:comprtos_runtime}
\end{figure}

This section describes CompartOS as a generic compartmentalization model and its design choices. The model is not
specific to a particular implementation but can be applied to mainstream embedded (operating) systems,
boot loaders, and programming languages. An overview of a CompartOS model system is shown in Figures~\ref{fig:comprtos_tooling}
and~\ref{fig:comprtos_runtime}. Both figures are further described in the design choices discussed next.

\begin{tcolorbox}
    CompartOS is a programming language- and linkage-based compartmentalization model.
\end{tcolorbox}

CompartOS relies on the underlying linkage model and format (e.g., ELF, Mach-O,
or PE) to define what a compartment is at compile time. The security policy between compartments is
specified at development and compile time and enforced at runtime load and link time.
Linkage-based compartmentalization has several benefits over other compartmentalization models:
\begin{itemize}
    \item Intuitive: Compartments are directly mapped to source-code files, objects, and
    libraries.
    \item Compatible: No need to reimplement or redesign existing software projects.
    \item OS-independent: No reliance on specific threading or process models, virtual memory
    support, or OS-enforced access control policies (e.g., filesystem-based ACL)
\end{itemize}

\begin{tcolorbox}
    A compartment is a linkage-based module.
\end{tcolorbox}
In CompartOS, the basis of a compartment is defined to be anything that could go into one or more
source-code files that will be compiled into a relocatable object module or a library. This
could optionally contain code (functions), data (variables), and other
sections. Compartmentalization is thus \textit{specified} at compile-time and \textit{enforced} at
the runtime linkage stage. That is,
CompartOS relies on the linkage model and modules to define the basis of compartments statically at
development and compile time.
Therefore, a compartment can encapsulate applications, libraries, device drivers, secret keys,
software updates, etc., independent of the task or process model.
Such benefits make it easy to compartmentalize projects like static baremetal
single-threaded ones, to rich \gls{EOS}es with multi-threading support, dynamic
memory, and filesystems.

\begin{tcolorbox}
    An inter-compartment security policy is defined by the language/linkage model.
\end{tcolorbox}

Since compartments are directly mapped to source-code files, it is natural to
make use of the programming language and linkage model to specify:
\begin{itemize}
    \item API: to define the visibility of each symbol to other compartments.
    \item Communication: to define the relationship with other compartments.
\end{itemize}

The specification is later on enforced using CompartOS and a CHERI-aware
secure loader.

\begin{tcolorbox}
    A compartment owns static linkage-based and dynamic resources.
\end{tcolorbox}

Besides statically defined linkage-based resources, a compartment can also own
dynamic resources it allocates at runtime. Furthermore,
compartments can own privileges to access IO memory regions and certain
privileged instructions, defined and granted as CHERI capabilities with
associated restricted permissions.

\begin{tcolorbox}
    Capability-based access control.
\end{tcolorbox}
In CompartOS, any access to resources has to be performed via a capability. Each
compartment has a capability table that contains CHERI capabilities to access linkage-based and
dynamic resources. The access control is specified at both language-based and OS/linkage-based
compartmentalization levels. A secure CHERI-aware compartment loader enforces both memory-safety
within a compartment (like in memory-safe languages) and spatial compartmentalization guarantees
between compartments like in capability-based microkernels.

\begin{tcolorbox}
    CompartOS provides a mechanism for compartmentalized fault isolation and recovery, not for catching specific security violation faults.
\end{tcolorbox}

While CompartOS is benefiting from CHERI's memory safety for catching security violations,
CHERI's security guarantees are not part of the model as a new contribution. That is, CompartOS does not
require or enforce a particular mechanism or policy for catching specific security violations
(e.g., buffer overflows, use-after-free, filesystem access control, etc.). It is
up to the embedded systems designer to deploy and implement mechanisms for catching security-related violations
and triggering architectural faults for them.


\begin{tcolorbox}
    A compartment does not trap to perform a protection domain switch.
\end{tcolorbox}
Unlike \gls{MMU} and MPU based compartmentalization techniques, CompartOS does not have the notions of
system calls or user, kernel, secure, unsecure, privileged or unprivileged. Compartment switches are normal function calls in the same privilege
ring and do not incur further microarchitectural overheads due to traps or hardware reconfiguration.

\begin{tcolorbox}
    No software monitors, hypervisors, or background checks.
\end{tcolorbox}
Unlike hypervisors, microkernels, and secure EOSes, CompartOS only requires a secure loader to
enforce the compartmentalization policy at load or boot time. Once compartmentalized, no further
background checks or monitoring happen in software. The integrity and confidentiality of the system
are maintained by virtue of capability-based access control and CHERI hardware.

\begin{tcolorbox}
    CompartOS' linkage-based compartmentalization can also support task-based compartmentalization.
\end{tcolorbox}
\label{sec:comp_models}

The CompartOS model could be used to support different types of compartmentalisation models:

\begin{itemize}
    \item \textbf{Task-based}: Single task  per compartment.
    \item \textbf{Library-based}: Multiple compartments per task.
    \item \textbf{Multitask-based}: Multiple tasks per compartment.
\end{itemize}

%

\begin{tcolorbox}
    All compartments execute in a single-address-space, MMU-less, single-privilege-ring environment.
\end{tcolorbox}

Except for CHERI, there is no reliance on specific hardware (e.g., \gls{MMU}, MPU, privilege rings) or OS
(virtualization, paging, address spaces, filesystem access control) features for protection or
privilege separation. CHERI enables enforcing software compartmentalization in a single flat
physical address space. Coarse-grained user and kernel separations are not required as CHERI could
isolate privileges for memory and system registers; thus, a single processor ring or privilege mode
is sufficient. This makes it possible for OSes, libraries, device drivers, and applications to be
all in the same privilege ring as different compartments.


\begin{tcolorbox}
    Dynamic runtime compartmentalization with actively malicious compartments.
\end{tcolorbox}

While formal verification, type-safe programming languages, certification, and static analysis tools provide
guarantees reasoning
about the security of a software system, adding new features, libraries, or
refactoring some
existing code requires considerable effort to maintain such guarantees.
Furthermore, static
guarantees are obtained against certain types of known models, vulnerabilities, and
exploits. Our
approach, on the other hand, allows for dynamic security, including creation,
isolation and
deletion of compartments during (boot, load, and unload) runtime. This helps protect against unknown
future vulnerabilities and exploits. Moreover, once deployed, we preserve the ability
to selectively
patch
and update specific components if needed without affecting the remaining system.

\section{Implementation}

In this section, we discuss how we implement the CompartOS model spanning three different software components:
a \textit{toolchain}, a \textit{dynamic loader}, and an \textit{embedded operating system}.
These parts collaborate to carry a compartmentalization policy from compilation and code generation through
to runtime using static and dynamic linkage of a set of modules.
We have extended the existing CHERI-LLVM toolchain, libdl dynamic loader, and the FreeRTOS EOS to implement the CompartOS model.
For simplicity, we refer to the implementations of those three software subsystems collectively as \textit{CheriFreeRTOS}.



%

\section{CHERI-LLVM: GPREL Addressing ABI}
\label{sec:llvm_gprel}

Current embedded operating systems, even the existing CHERI-based ones, compile and
build statically and get linked into a single binary image that executes
in a single address space and single privilege ring.
This means that all code can call any other code, and access any other data.
Global variables and functions can thus be referenced from anywhere.
Accesses happen in a Program Counter Relative (PCREL) manner, and in the case of CHERI,
only one capability table is shared for the entire image.

\begin{figure}
    \centering
    \includegraphics[width=\columnwidth,height=7cm]{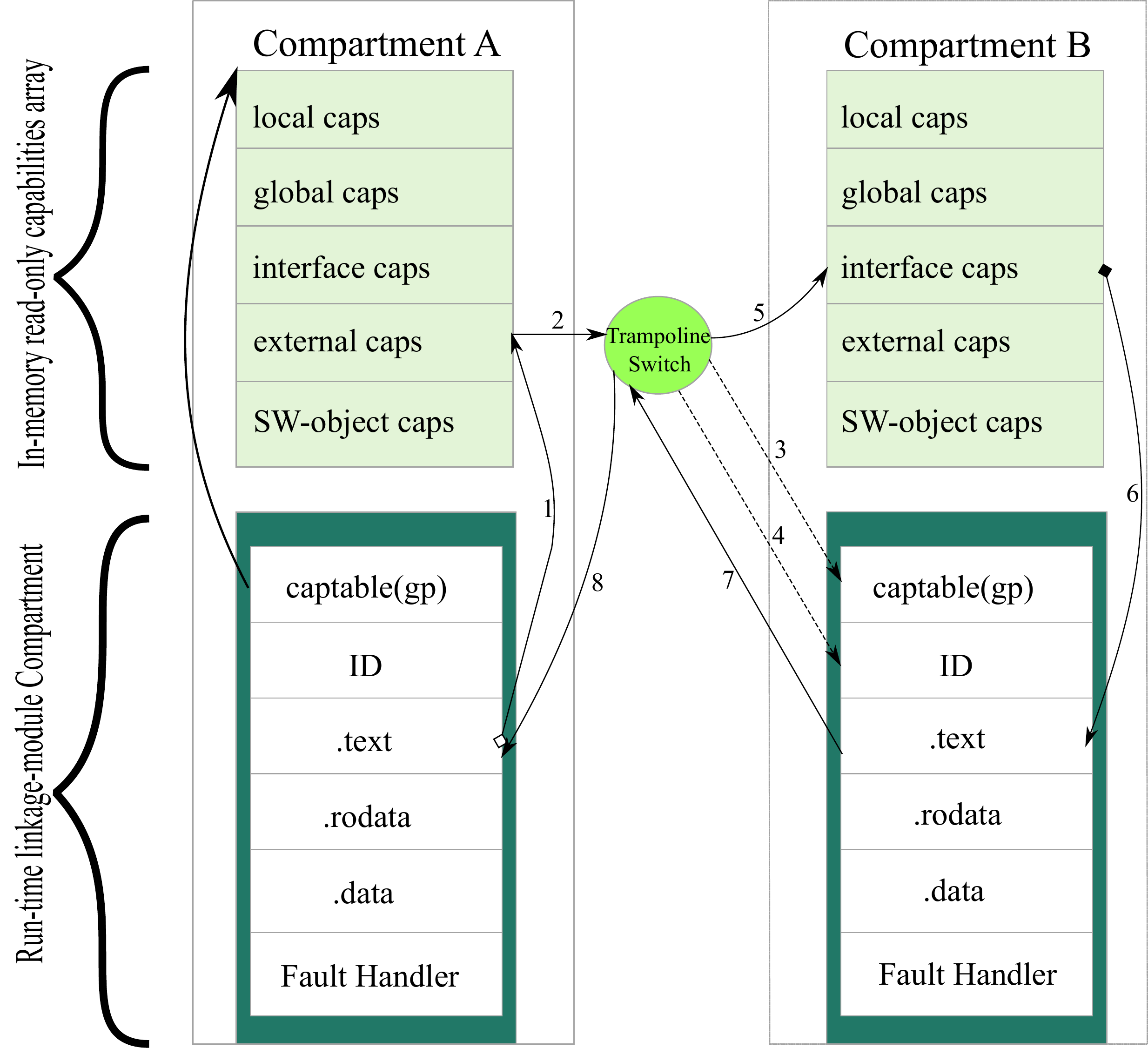}
    \caption{A runtime example of a compartmentalized system performing domain switching. 
        The switch starts with a
        function call that references a capability from the externals capability list (\#1) which
        points to a small, read-only trampoline (\#2) that performs
        compartment switches by setting the new captable (\#3)
        and compartment ID (\#4) after storing the caller's context. It then jumps via the
        interface capability (\#5) provided by the target compartment.
        This interface capability points to the function within its associated compartment (\#6).
        Upon its return (\#7),
        the trampoline restores the caller's context, captable, and ID, then returns back to
        the caller function (\#8).
    }
    \label{fig:comp_illustration}
\end{figure}

Having a single globally shared captable that is PCREL-accessed suffices for UNIX-based systems with well-defined process
models and \gls{MMU}s (such as in CheriABI~\cite{davis2019cheriabi}), but is inadequate to support compartmentalization in embedded systems with a
single physical memory space.
For example, if this PCREL addressing was employed in EOSes
with CompartOS, it would create a single captable granting all compartments (e.g.,
applications, drivers, kernels, libraries) unrestricted shared  access to all globals.
CompartOS, therefore, requires runtime support for a single-address-space image with multiple independent captables
and bounded functions (e.g., that cannot address anything relative to their address, except their code)
as shown later in Figure~\ref{fig:comp_illustration}.

We overcame this challenge by adding new support for per-compartment captables; however, this required changes to the
ABI.
Specifically, we used another architectural capability register, the
Capability Global Pointer (CGP) and changed the compiler ABI for compartments to
generate GP-relative (GPREL) instead of PCREL accesses. Each
compartment's CGP holds its root captable.

%

\section{libdl: Secure Dynamic Linker and Loader}
\label{cherilibdl}

The original \texttt{libdl} (a dynamic linker and loader library)
provides no protection between linkage modules such as objects, libraries and the
base ELF image (e.g., the OS kernel), and in fact, has no mechanism it could even use to do so.
That is, once new objects or libraries are loaded, the resulting system is
indistinguishable from a baseline system statically linked at build-time.
In runtime, object modules can call global
external functions in other modules. This is considered the baseline for a dynamically loaded
system by libdl. Even if we build libdl and its loaded modules in CHERI C,
there will only provide spatial pointer memory-safety, but not  isolation and compartmentalization
between loaded linkage modules, which can still access a globally shared captable
between modules.
We have extended \texttt{libdl} to support CHERI and CompartOS linkage-based software compartmentalization.
\texttt{libdl} acts as a secure boot loader or a trusted firmware. That is, it is the most privileged
piece of software that first takes control of the hardware on reset, holds access to the CHERI root capabilities (i.e., initial capabilities
that hold all privileges such as PCC and DDC), and then loads
and grants compartments the minimum restricted capabilities they need to perform their job.


%




\subsection{Compartment Creation}


A compartment is dynamically created by calling \textit{dlopen()} with the path to the module in the
file system. After
\texttt{libdl} allocates the memory necessary for loading and linking, it also allocates
capability table with an initial size of the defined symbols count
within that module.
External function symbols are treated as inter-compartment domain switches.
Those are detected by
\texttt{libdl} during a compartment creation.
For every external function call within a compartment, an
additional lightweight function trampoline is emitted (discussed later).
Capabilities can be minted (by libdl)
from other capabilities through monotonic action. Minting means deriving a new capability
out of an original one, with the same or less permissions.
If an inter-compartment call is allowed (by some user-defined policy),
\texttt{libdl} mints a
capability from the callee’s captable and installs it in the caller’s captable
dynamically.
The captable will come to hold capabilities
for four different classes of linkage-based symbol capabilities: local, global, interface, and external.
Interface capabilities form a compartment's API that other compartments can have minted external capabilities to.




On boot time, \texttt{libdl} automatically loads and links all the compartments that the embedded system designer
designates in a configuration file. 
The configuration file is considered part of the security policy.
CheriFreeRTOS aligns with C and ELF linkage-based symbols attributes. For example, it creates read-only capabilities
for code and constants. Similarly, stack capabilities and \textit{malloced} memory (from
compartments) do not have execute permissions.

\subsection{Compartment Switch}
\label{sec:comp_switch}

A compartment switch triggers a protection domain switch between distrusting compartments,
as shown in Figure~\ref{fig:comp_illustration}.
\texttt{libdl} detects inter-compartment calls at load time and emits a read-only, \textit{Sentry} trampoline that
performs a compartment domain switch accordingly. \textit{Sentry} (sealed entry) functions are CHERI capabilities
that can only be jumped to but not read from or written to.
Additionally, \texttt{libdl} wraps function
pointers with a trampoline since those can be leaked between compartments.
The trampoline consists of metadata and code to define which compartment a function belongs to, the
destination compartment captable, and a bounded capability to the function
itself (see Listing~\ref{lst:tramp_code}). \texttt{libdl} sets up the metadata (function capability,
destination captable, and compartment identifier) during compartment loading.
The only privilege a trampoline has is the metadata.


\begin{lstlisting}[float,floatplacement=H,basicstyle=\tiny,caption={Compartment trampoline pseudo
        code},label={lst:tramp_code},frame=single,captionpos=b]
    xPortCompartmentEnterTrampoline:
    % Metadata
    .Lfunc:      .zero CAP_SIZE % Callee's function capability
    .Lcaptable:  .zero CAP_SIZE % Callee's capability table
    .Lcompid:    .zero CAP_SIZE % Callee's compartment ID
    // Compartment/Domain Switch
    Save caller's context (register set, GP, compid)
    Setup new GP/captable
    Set currentCompartmentID = Lcompid
    (Optional) Restrict/bound the callee's stack
    Call the destination function
    Restore caller's context (register set, GP, compid)
    Return to the caller compartment
\end{lstlisting}

\section{FreeRTOS: an Embedded OS}
\label{sec:cherifreertos}
In this section, we describe the subsystems where the dynamic linker and an embedded OS may need
to cooperate. The required change in an EOS to support linkage-based software
compartmentalization is its threading support. Optionally, if the EOS supports dynamic
memory allocation, some changes might also be required there.
We have chosen FreeRTOS as an EOS, which is widely deployed and is representative.


\subsection{Compartments and Threading}
In CheriFreeRTOS, a thread can enter exactly one compartment at a time, and compartment entrances can be nested in a
function-call manner. A compartment is also re-entrant from different threads, assuming the
function is re-entrant and thread-safe. When a thread enters a
compartment, it donates its remaining stack to the compartment.
The bookkeeping is handled by the FreeRTOS \texttt{TCB\_t} structure, which is
extended to contain a compartment context, including the current compartment ID
(xCompID), the caller’s return address, and the caller’s return stack.



\subsection{Software Fault Handling}
\label{software_fault_handling}

We have modified FreeRTOS to support handling architectural exceptions for the RISC-V port.
The baseline port does not handle or support any form of exception handling and instead
hangs or performs a reset instead.

%

\subsubsection{Fault Handler Registration}
There could be different types of fault handlers in the CompartOS model, discussed in Section~\ref{sec:eval_fault}.
Those can involve either libdl, FreeRTOS or both of them. In CheriFreeRTOS, a custom fault handler can be
registered per compartment by defining and implementing a \textit{CheriFreeRTOS\_FaultHandler} function
for each compartment at development and build time. During the loading process, libdl searches for any occurrence of
\textit{CheriFreeRTOS\_FaultHandler} as a function symbol and registers it as a callback
function for the loaded compartment. Otherwise, if libdl could not find \textit{CheriFreeRTOS\_FaultHandler},
it applies default fault handling techniques.


\subsubsection{Runtime Fault Handling}
If an exception occurs, the modified FreeRTOS architectural exception handler checks if the current task is
running in a compartment (by checking the associated per-task compartment structure),
and if so, it jumps to a registered custom per-compartment software fault handler (if provided).
In the case of Return-Error and Compartment-Kill (see Section~\ref{sec:eval_fault}), the exception handler returns back to the trampoline that performed the
compartment switch, which returns to the caller compartment with an optional return error code.
These per-compartment software fault handlers improve fault tolerance and maximize availability.

\section{Evaluation}
\label{section:evaluation}

We evaluate CompartOS through our CheriFreeRTOS prototypes with regards to performance,
compatibility, availability, security, and practicality,
demonstrating fully-functional real-world use cases.
We ported and developed MPU and CHERI-aware CheriFreeRTOS to run on
different variants of RISC-V processors on the VCU-118 Xilinx \gls{FPGA} board.
The (CHERI-)RISC-V processor's HDL and CheriFreeRTOS software stacks are open-source.
This section starts with measuring the hardware costs in terms of \gls{FPGA}
LUTs for unprotected, MPU/\gls{PMP} (the RISC-V analogous version of Arm's MPUs), MMU, and CHERI processors. Next, we evaluate different
security models implemented by our
software variants in CheriFreeRTOS, including MPU, CHERI/PURECAP, and CompartOS through
a set of micro- and macro- benchmarks.

We require the hardware to provide at least a few
megabytes of RAM memory to support a filesystem, compartmentalization, and
dynamic loading and linking. Thus, we exclude memory resource
restrictions, as we target high-end embedded system applications.

\subsection{Hardware Measurements}
\label{sec:hardware_eval}

We use an \gls{FPGA} softcore called Flute by Bluespec which is a RISC-V processor and is comparable to Armv7
low-end cores (e.g., A9) that are commonly used in high-end embedded systems. We synthesize Flute
to run at 100 MHz with 5 pipeline stages, 32-bit mode, without a Floating Point Unit (FPU).
Flute's privilege modes are similar to Arm's. For example, Machine-mode (EL3) is used for
firmware, embedded systems or \gls{RTOS}es, Supervisor-mode (EL1) runs conventional OSes that support
paging and virtualization. Flute
is also extended to support CHERI-RISC-V~\cite{cheririscv2022}. We have built four different processors of Flute for evaluation,
as shown in Table~\ref{tab:fluteprotprocconfig}. The hardware consumption numbers are shown in Table~\ref{tab:fluteppa}.
We provide these numbers just for completeness with a notice that we mostly
rule out the area and power requirements as we target high-end embedded processors.
Further, the power row in the table represents the approximate static power consumption, but that might not be a good
indication of the runtime power consumption as that depends on the software workload and IO operations (e.g.,
DRAM and peripherals access).
We find that the \textbf{PMP} variant with 16 regions
adds 7.6\% LUTs overhead compared to  \textbf{NOPROT}, while the
\textbf{MMU\_PMP} one takes slightly more. \textbf{CHERI} takes the most LUTs overhead of 10.4\% as
it implements a
rich capability-based ISA. On the other hand, \textbf{CHERI} takes the least Registers area
overhead of
1.3\% compared to 2\% and 2.5\% overheads for \textbf{PMP} and \textbf{MMU\_PMP}, respectively, as
\textbf{CHERI} does only run
in Machine-mode and thus requires no further Supervisor/User privilege mode registers, \gls{TLB}s, or \gls{PMP}s.
We think that such overheads are low for the security benefits that CHERI and CompartOS
bring.


\begin{table}[t]
    \centering
    \begin{adjustbox}{max width=\columnwidth}
        \begin{tabular}{lllll}
            \toprule
            & \textbf{NOPROT}      & \textbf{PMP}   & \textbf{MMU\_PMP} & \textbf{CHERI}  \\
            \midrule
            RISC-V Privilege Modes & M              & M, U          & M, S, U      & M      \\
            Protection             & N/A            & PMP (MPU)     & PMP, MMU     & CHERI  \\
            PMP regions            & N/A            & 16            & 16           & N/A     \\
            TLB                    & N/A            & N/A           & 16 entry direct-mapped & NA \\
            \bottomrule
        \end{tabular}
    \end{adjustbox}
    \caption{Hardware configurations of different Bluespec processors we build for evaluation.}
    \label{tab:fluteprotprocconfig}
\end{table}


\begin{table}[t]
    \centering
    \begin{adjustbox}{max width=\columnwidth}
        \begin{tabular}{lllll}
            \toprule
            & \textbf{NOPROT}     & \textbf{PMP} & \textbf{MMU\_PMP}     &
            \textbf{CHERI}  \\
            \midrule
            LUTs                 & 92222          & 99810 (7.6\%)         & 101132 (8.8\%)       & 102955 (10.4\%) \\
            Registers            & 119012         & 121498 (2.0\%)       & 122096 (2.5\%)      & 120600 (1.3\%) \\
            Power (W)            & 0.212          & 0.213 (.47\%)        & 0.221 (4.1\%)       & 0.245 (13.5\%)\\
            \bottomrule
        \end{tabular}
    \end{adjustbox}
    \caption{Power and area results for different variants of Flute. Overhead percentage is calculated against NOPROT.}
    \label{tab:fluteppa}
\end{table}


\subsection{Software Variants}
We have evaluated multiple FreeRTOS-based benchmarks and case studies on the previously discussed
hardware variants and the following software setups that we developed:

\begin{itemize}
    \item \textbf{INSECURE-STATIC:} Statically-linked, unmodified (and unprotected) baseline RISC-V
    software built without CHERI, \gls{PMP}, or \gls{MMU} support.
    \item \textbf{INSECURE-DYNAMIC:} Same as \textbf{INSECURE-STATIC}, but supports dynamically loading and linking
    RISC-V modules using \texttt{libdl}. This is running on the \textbf{NOPROT} hardware processor.
    \item \textbf{PMP-4-TASKS:} Statically-linked RISC-V software providing coarse-grained
    task-based and MPU-based compartmentalization (4 regions). This is running on the \textbf{PMP} hardware variant.
    This is similar to FreeRTOS-MPU, TockOS, and RTEMS' MPU-based security models.

    \item \textbf{PMP-N-OBJS:} Automatic linkage-based compartmentalization where every object
    module
    (source file) is a compartment protecting N number of resources and protection is enforced
    using \gls{PMP}/MPU, and is dynamically loaded and linked. This is running on the the \textbf{PMP}
    hardware processor. This is similar to ACES, TyTan/TrustLite, and uVisor's security models.

    \item \textbf{PMP-N-LIBS:} Like \textbf{PMP-N-OBJ}, but compartments are library modules rather
    than object modules.
    \item \textbf{PURECAP:} Statically-linked pure-capability CHERI-RISC-V software providing
    complete (spatial) pointer safety, but no compartmentalization, running on the the \textbf{CHERI}
    hardware processor. This is similar to task-based CheriRTOS model.
    \item \textbf{COMPARTOS-OBJS:} Our new automatic linkage-based compartmentalization where every source
    file (object module) is a compartment, protection is enforced using CHERI, and is dynamically
    loaded and linked. This is running on the \textbf{CHERI} hardware processor.
    \item \textbf{COMPARTOS-LIBS:} Like \textbf{COMPARTOS-OBJ}, but compartments are library
    modules rather than object modules.
\end{itemize}

\subsection{Microbenchmarks}
\label{sec:microbenchmarks}

We have developed a custom microbenchmark from scratch to measure the performance of
the lowest level, most critical and frequently executed code paths in embedded OSes
in the different software variants that we discussed.
There are two main communicating compartments: a \textit{sender} and a \textit{receiver}.
Communication is done over function calls, compartment calls/switches,
and \gls{IPC} message-passing.
In the object-based compartmentalized variants (e.g., COMPARTOS-OBJS and PMP-N-OBJS),
each compartment is an object file. The sender compartment has 18 resources to protect,
and the receiver compartment has 41. Those are memory regions representing pointers,
global variables, UART regions, and functions.

\subsubsection{Task-based IPC Evaluation}

\begin{figure}
    \centering
    \includegraphics[width=\columnwidth]{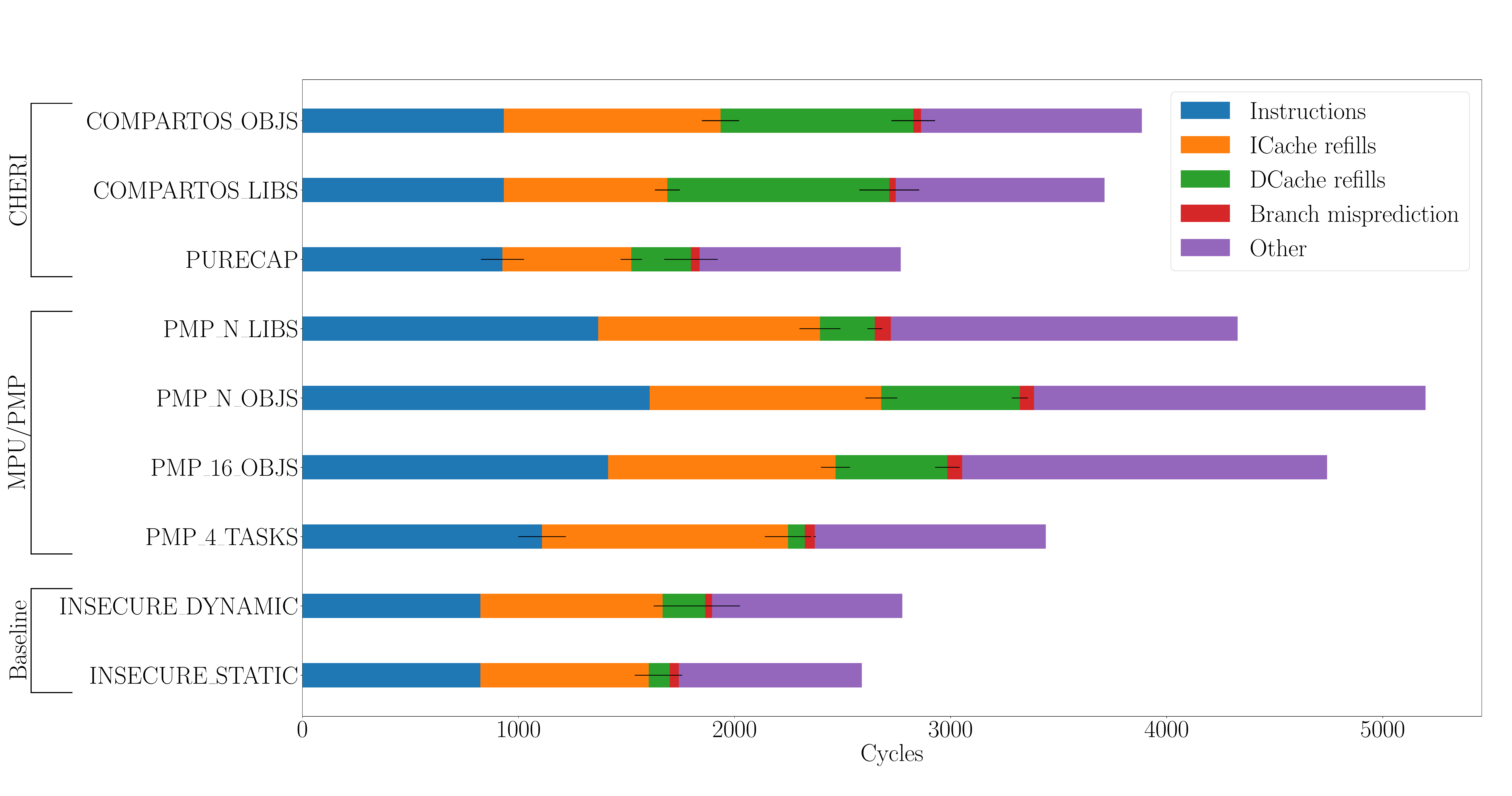}
    \caption{The cost, in cycles, of task-based domain switching, sending one byte
        from one task to another.}
    \label{fig:ipc_queues_abs}
\end{figure}

%

In secure EOSes (like in seL4, FreeRTOS-MPU, and TockOS), secure domain switches
are often task-based using IPC and message passing. In
Figure~\ref{fig:ipc_queues_abs}, we show the
performance overhead of performing task-based domain switches, sending 1 byte over a FreeRTOS queue.
Some instructions are multi-cycle and/or require pipeline flushes; those are placed in the \textit{Other} segment. Other
instructions take one cycle, and they are accounted for in the \textit{Instructions} segment. The
remaining segments are cache misses and branch misprediction latencies.
The PMP variants take the most instructions represented in \textit{Instructions} and \textit{Other}
segments as they have to perform system calls, reconfigure PMP registers (see Figure~\ref{fig:domain_switch_instrs})
that require loading
permissions from memory and writing the PMP system registers. CompartOS variants do incur negligible
overhead in data cache misses, likely because of the partitioned capability table per compartment,
unlike PURECAP.

\subsubsection{Performance Evaluation of Domain Switching}
\label{sec:domain_switch_eval}

\begin{figure}[]
    \centering
    \includegraphics[width=\columnwidth]{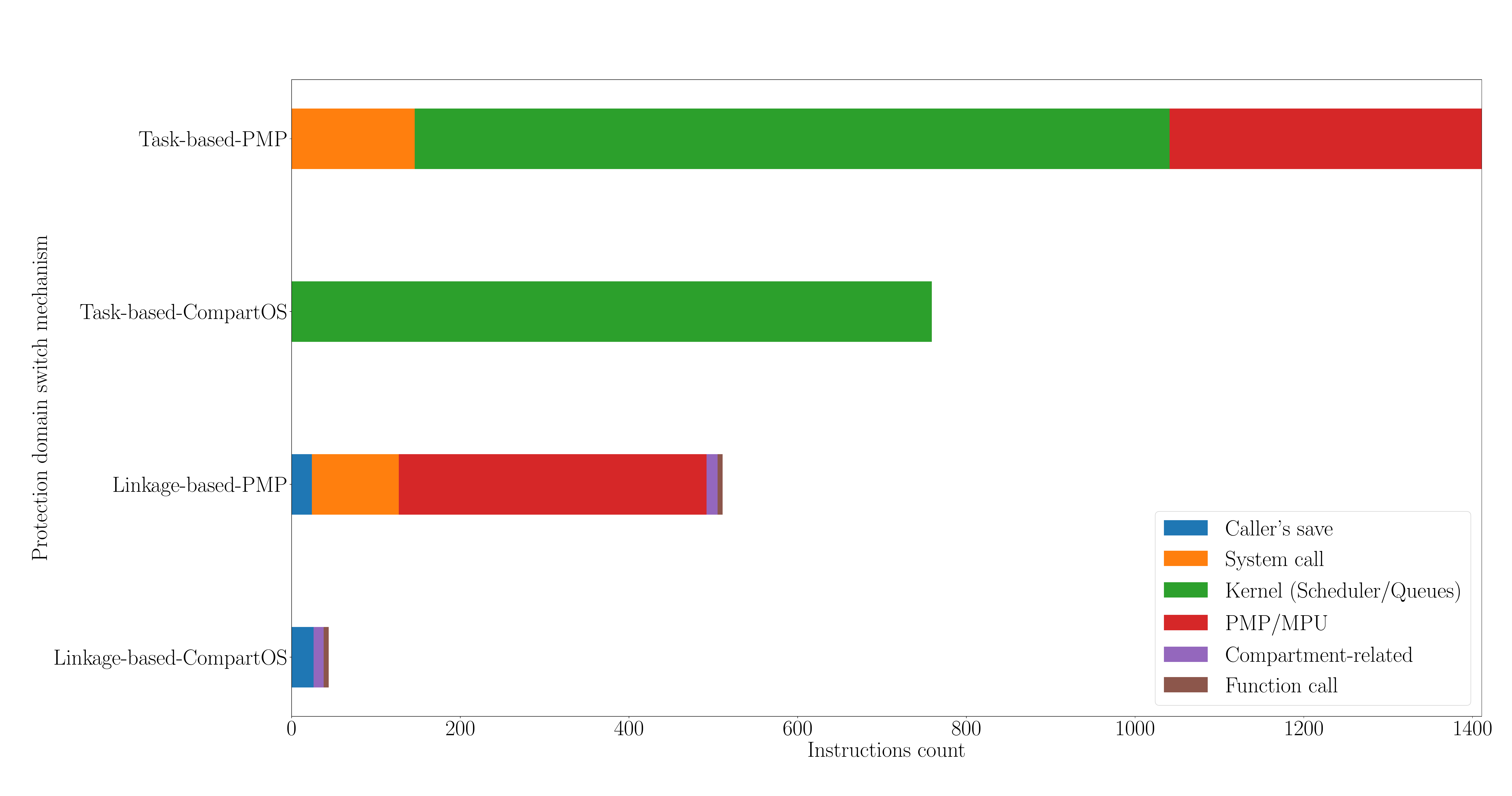}
    \caption{Comparison between different protection domain switch mechanisms cost in instructions count.}
    \label{fig:domain_switch_instrs}
    \includegraphics[width=\columnwidth]{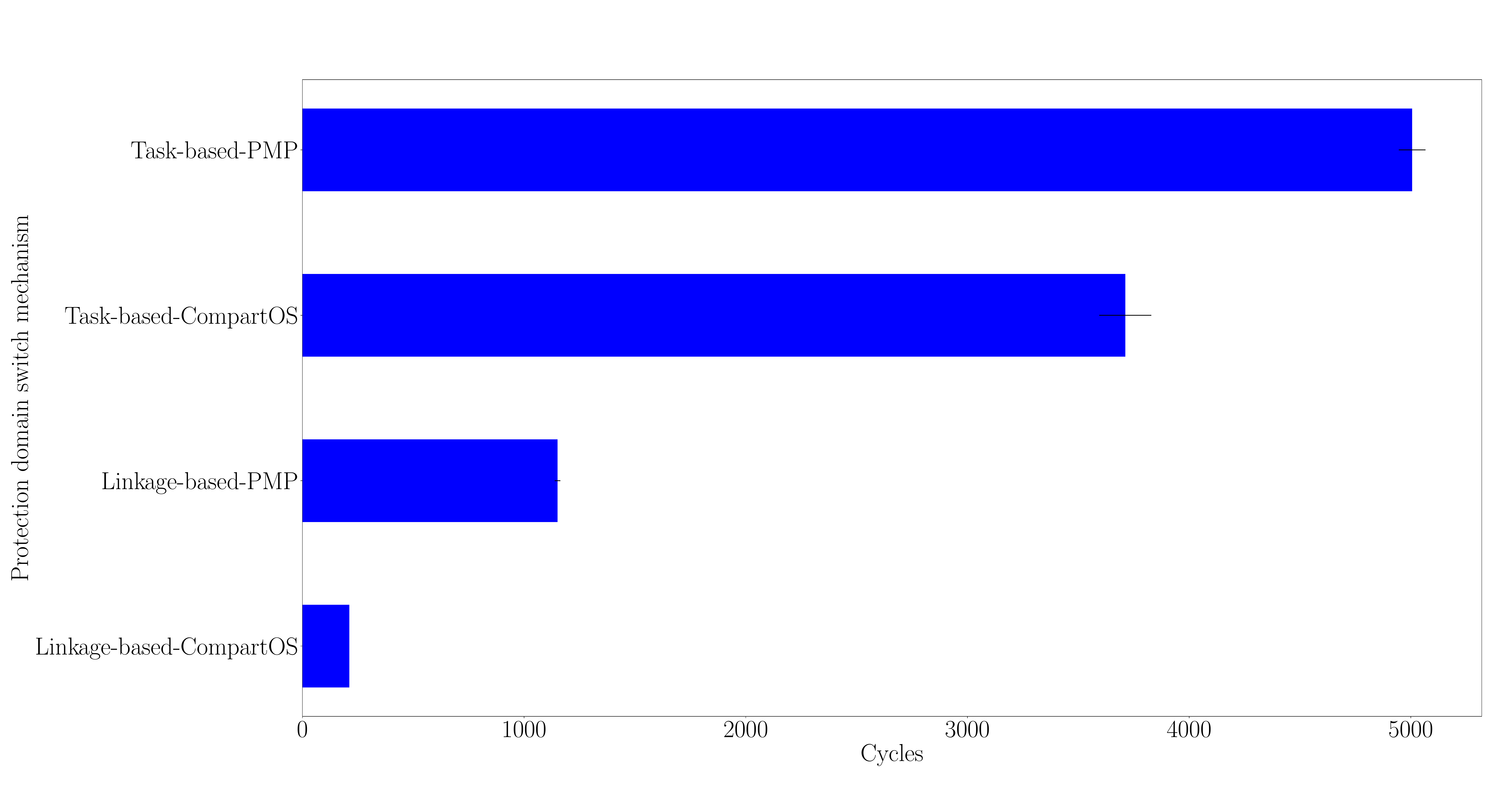}
    \caption{Comparison between different protection domain switch mechanisms cost in cycles count.}
    \label{fig:domain_switch_cycles}
\end{figure}

We compare CompartOS' linkage-based domain switch against state-of-the-art PMP/MPU task-based
domain switch in Figures~\ref{fig:domain_switch_instrs} and~\ref{fig:domain_switch_cycles}.

Unlike task-based, linkage-based compartmentalization performs compartments switches during function calls.
Looking at Figure~\ref{fig:domain_switch_instrs}, we notice that the Task-based-PMP instructions are consumed
in performing system calls, IPC/kernel path, and PMP configurations. On the other hand, CompartOS'
task-based compartmentalization does not
require any system calls or PMP reloads, and compartment switches happen implicitly during task context
switches when the CGP register is swapped.
Based on Figure~\ref{fig:domain_switch_cycles}, we conclude that \textit{CompartOS' task-based domain crossing is 26.6\% faster than MPU-based,
task-based, IPC mode, implemented by state-of-the-art deployed systems (like in FreeRTOS-MPU and TockOS)}.

Similarly, we compare CompartOS linkage-based domain switch against the state-of-the-art MPU-based
linkage-based systems (e.g., ACES, uVisor, TrustLite/TyTan). The PMP reconfigurations and system calls dominate
the cost of Linkage-based-PMP; both are not required in CompartOS.
\textit{We conclude that CompartOS' linkage-based domain switch is 85\% faster than the
    most similar MPU-based compartmentalization state-of-the-art systems.}

Finally, we compare CompartOS' linkage-based compartmentalization against off-the-shelf deployed
OSes that rely on task-based, MPU-based compartmentalization.
CompartOS offers intra-thread domain switches thanks to its linkage-based (rather than task-based)
design. We conclude that
\textit{CompartOS' linkage-based domain switch is 95\% faster than the
    off-the-shelf message-passing IPC in task-based, MPU-based models implemented by state-of-the-art deployed systems (e.g., FreeRTOS-MPU and TockOS).}

\subsection{TCP/IP Benchmarks}
\label{sec:tcpip_bench}

We perform TCP/IP evaluation to understand how a real-world mainstream system would perform across the
different protection models implemented by our FreeRTOS software variants.
The evaluation includes response time, throughput, and security.
As most feature-rich \gls{IoT} systems do have a TCP/IP stack
for communication and connectivity, this TCP/IP use case is the most critical, complex,
and realistic application we are discussing so far. It is important to note that we evaluate
mainstream TCP/IP centric applications that are deployed by FreeRTOS and are published publicly~\cite{freertostcpipdemos22},
including multiple internet servers and protocol stacks built as library compartments. Those are over 9 library compartments
(in COMPARTOS-LIBS) and over than 50 compartments (in COMPARTOS-OBJS and PMP-N-OBJS),
each with hundreds of resources, and over than 10 compartment switches in the FTP use case, discussed later. Hence, scalablity
is evaluated as well.
Further, the compartments are mutually separate, often distrusting, and sometimes come from different vendors
with different criticalities and real-world published vulnerabilities.
We have not developed any of the code-based (which is about 100 KLoC) ourselves,
and, thus, we also evaluate the compatibility and applicability advantage of CompartOS.
The FPGA board running the benchmarks communicates with a host computer and are connected over a 1 Gigabit Ethernet (cross-over).

%
%
%
%

\subsubsection{Response Time}

\begin{figure}
    \centering
    \includegraphics[width=\columnwidth]{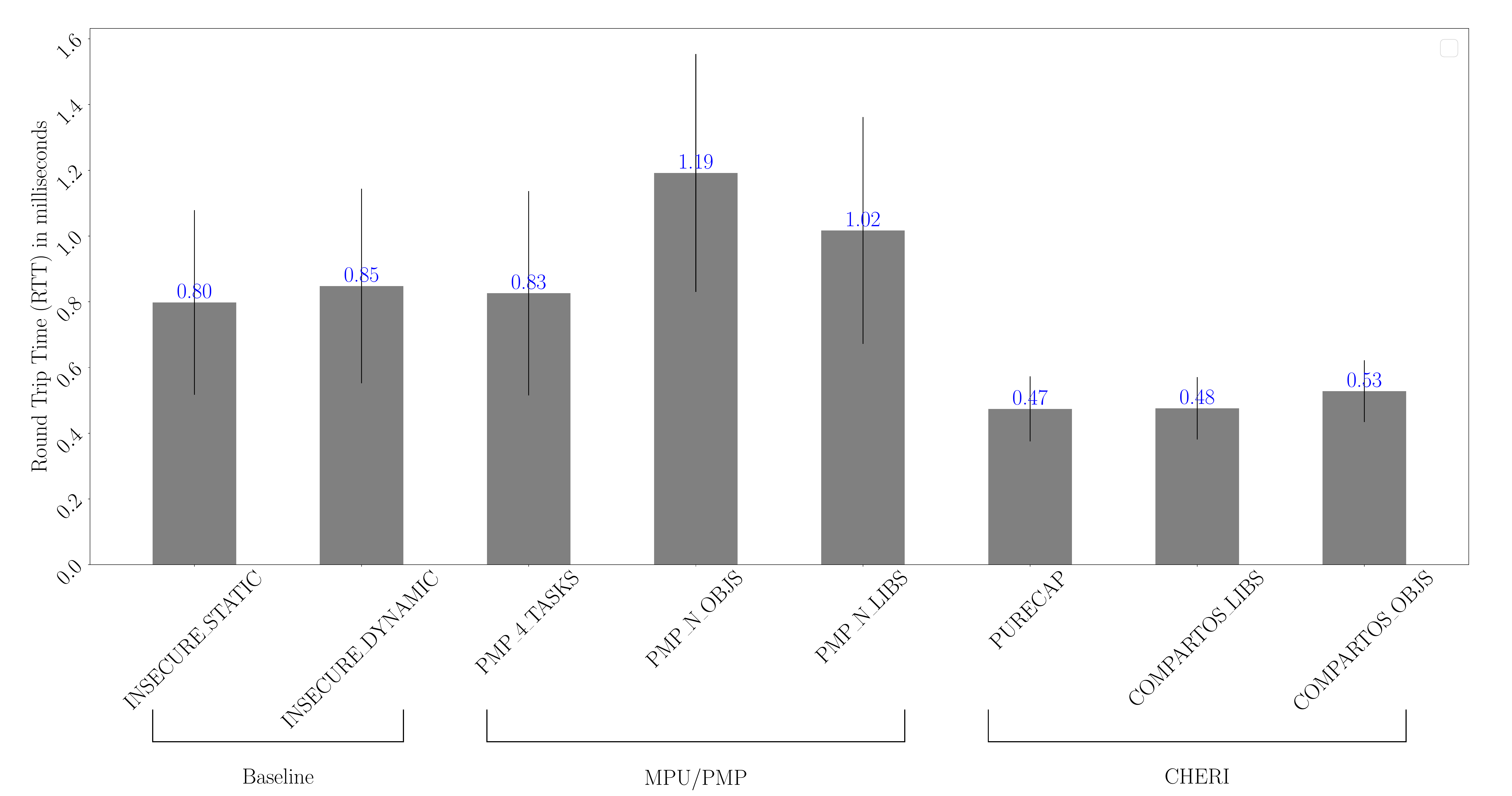}
    \caption{Small packet round-trip latency with ping (lower is better).}
    \label{fig:ping_performance}
\end{figure}

We perform response time evaluation by sending a total number of 100 ping packets and
getting the average and standard deviation numbers reported by the ping command on a Linux PC.
The command calculates the standard deviation of the Round Trip Time (RTT), which is
shown in Figure~\ref{fig:ping_performance}.
Ping sends a single ICMP packet from the client to the server and receives back an acknowledgment packet.
This makes the code path short and fast and thus maximizes the overheads. Further, the ICMP
packet processing only incurs one domain switch in the case of library compartmentalization from the
FreeRTOS ISR to the TCP/IP stack. The \textbf{PMP-4-TASKS}
is also effectively doing nothing when it comes to protection; FreeRTOS-MPU does not try to rework
the TCP/IP stack (being relatively complex and large) to use task-based MPU compartmentalization as that is
likely inadequate due to the MPU limitations mentioned so far.

\textbf{PURECAP} is 62.66\% faster \textbf{INSECURE-STATIC} baseline,
because CHERI doubles the register sizes, which halves the number of
instructions spent in \textit{memcpy} that is frequently used in the TCP/IP code paths.
We cannot use a non-CHERI-aware \textit{memcpy} in this case, as events
and the IPC buffers include pointers.
However, in Section~\ref{section:tcpip_throughput}, we amortize this
advantage. 
Comparing he\textbf{ COMPARTOS-LIBS} against \textbf{PURECAP} gets us the compartmentalization and
protection domain switches overhead. This is a negligible 3.6\% for such a small operation.
Similarly, if we compare \textbf{PMP-N-LIBS} against \textbf{INSECURE-STATIC}, we find
that PMP-based compartmentalization adds 25\% overhead.
A large part of this high overhead is because of the domain
switch between FreeRTOS ISR and the TCP/IP stack, which is dominated by costly PMP reconfigurations.

\subsubsection{Throughput}
\label{section:tcpip_throughput}
We measure the throughput of different implemented models by uploading
a file from the host computer to an FTP server running on FreeRTOS.
This is the most realistic workload and resembles a real-world use case where there are
different protocols, compartments, and subsystems triggering multiple protection domain
switches. The aim is to investigate how different implementations of protection models affect the overall
performance by amortizing the cost of timer interrupts, \textit{memcpy} advantage in CHERI, and single-packet
processing time, discussed in the previous section. We replaced all of the \textit{memcpy} calls that do not
contain pointers to use 4-byte integer registers copies instead of 8-byte CHERI registers,
across all variants, including PURECAP and COMPARTOS.
For each received packet, there are multiple protection domain switches between
task-based compartments, multi-threaded compartments, and library-based compartments.
For instance, a packet is received by the FreeRTOS ISR, which jumps to the VCU-118 network
device driver, which is part of a multi-task compartment (\textit{freertos\_tcpip}) as it contains
a thread for the ISR and a thread for the TCP/IP stack processing. Communication between
them is IPC-based over queues, thus triggering a context switch that implicitly performs
a protection domain switch. Later on, other compartments like \textit{tcp\_servers} and
\textit{ftp\_server} and the filesystem are called to process the FTP commands and write
the file before sending back an acknowledgment and response packet to the host.
Overall, the processing includes multiple domain switches, IO handling, memory-intensive
operations, and per-packet and scheduling handling, thus demonstrating a rich and real
diverse workload that is commonly used in deployed embedded systems.

\begin{figure}
    \centering
    \includegraphics[width=\columnwidth]{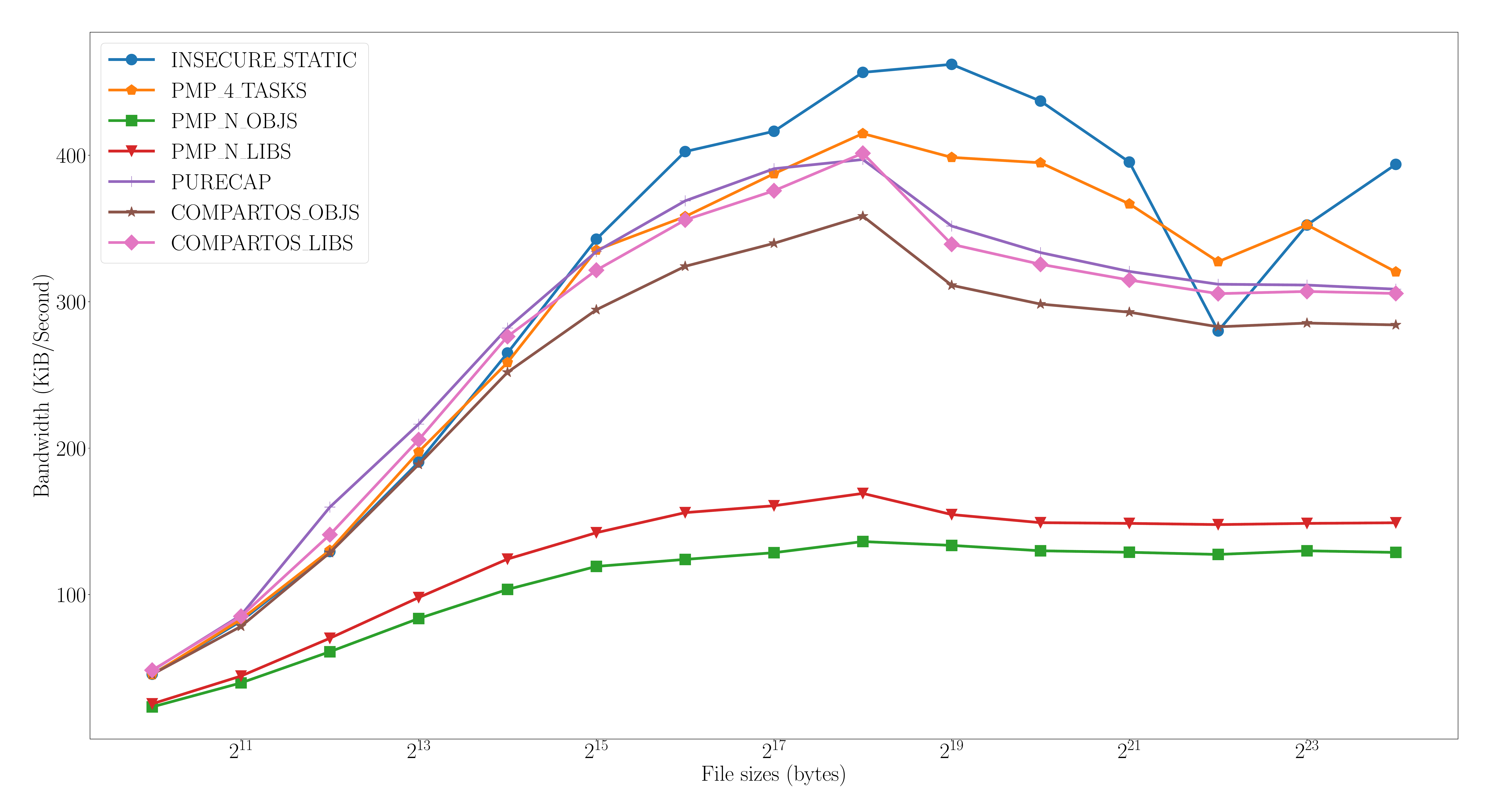}
    \caption{FTP upload bandwidth of varying file sizes.}
    \label{fig:ftp_throughput_varied_noncheri_memcpy}
\end{figure}

Figure~\ref{fig:ftp_throughput_varied_noncheri_memcpy} shows the absolute bandwidth of uploading
different file sizes from the host to FreeRTOS over FTP. The most relevant and important
variants there are \textbf{COMPARTOS-LIBS} and \textbf{PMP-N-LIBS}, with library compartments represented.
The bandwidth stabilizes after a certain file size
as the host PC cannot send packets any faster due to its hardware and Linux networking subsystems.
Figure~\ref{fig:ftp_throughput_varied_noncheri_memcpy} shows that the linkage-based PMP variants
are quite slow compared to all other variants, including CompartOS and PURECAP, regardless
of the file size. The bandwidth overheads further increase across the PMP variants
when the file size increases. This suggests that PMP-based compartmentalization is not
as scalable as CompartOS.

\begin{figure}
    \centering
    \includegraphics[width=\columnwidth]{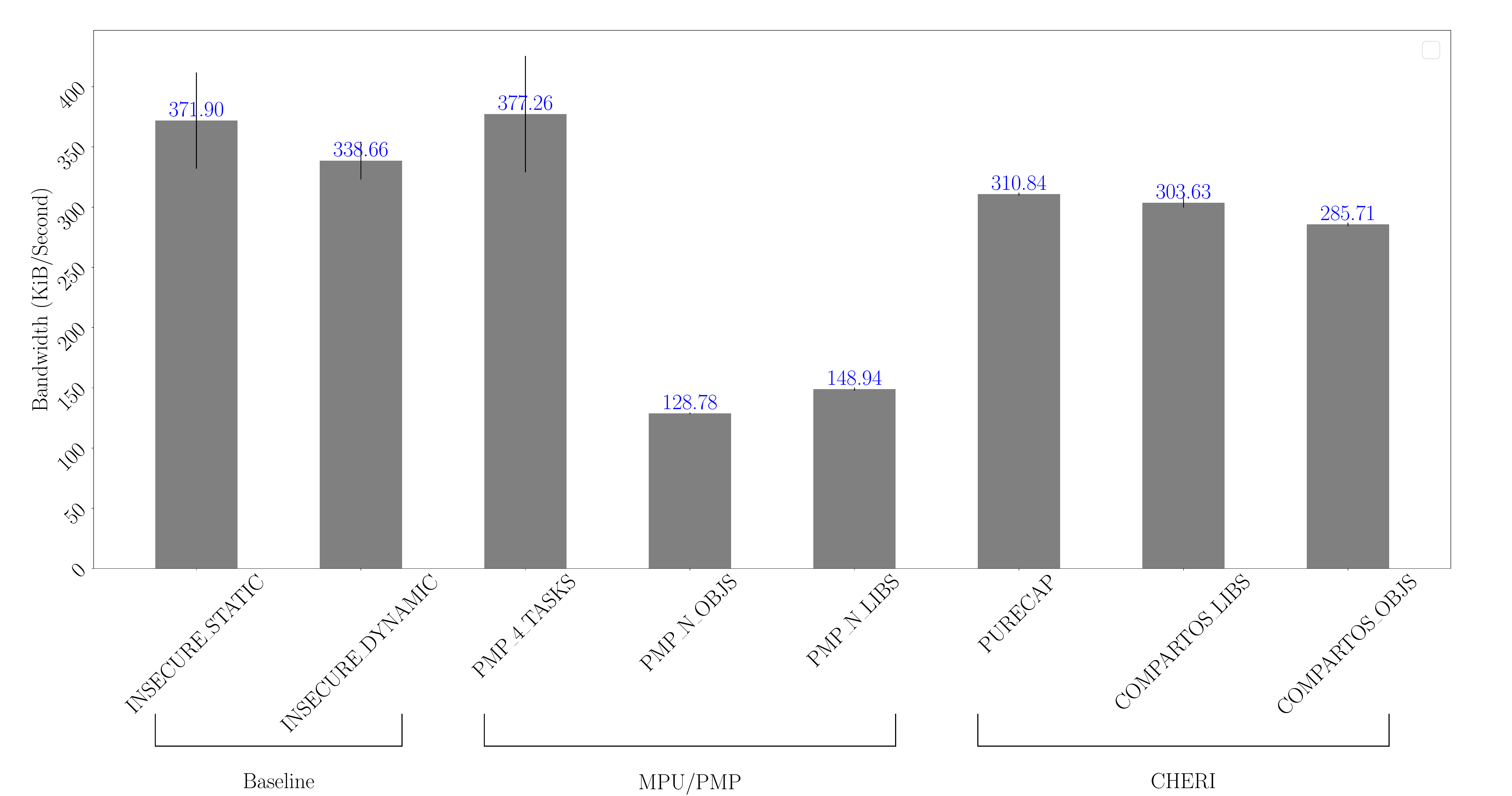}
    \caption{FTP upload bandwidth for 8 MiB file size.}
    \label{fig:ftp_throughput_fixed_noncheri_memcpy}
\end{figure}
%

In Figure~\ref{fig:ftp_throughput_fixed_noncheri_memcpy}, we pick a file size of 8 MiB
when the bandwidth stabilizes and compare the overheads against \textbf{INSECURE-STATIC}.
The \textbf{PMP-N-LIBS}
has a significant 60\% overhead while \textbf{COMPATOS-LIBS} only incurs 18\% overhead.
Further, \textbf{COMPATOS-LIBS}'s overhead is less than 2\% higher compared to \textbf{PURECAP}'s,
which suggests that the compartmentalization overhead (GPREL addressing, trampolines, and per-compartment captables)
is negligible. Finally, we conclude that\textit{ CompartOS' TCP/IP bandwidth is 52\% faster than linkage-based,
MPU-based compartmentalization.}

\subsubsection{Security}
\label{section:security}

We evaluate the security aspects of the new compartmentalization model by analyzing the integrity,
confidentiality, and availability of a compartmentalized FreeRTOS vulnerable system.
This yields a good speculative indication about the nature and impact of future unknown
vulnerabilities, and how they can be mitigated by CompartOS.
We use the National Vulnerability Database (NVD)~\cite{freertosnvdcves} from the US government's
National Institute of Standards and Technology (NIST) 
Based on our analysis, we found out that at least 10 out of
13 FreeRTOS CVEs~\cite{freertoscves} would have
been caught directly by CHERI, protecting against critical integrity and confidentiality vulnerabilities.
CompartOS takes security further to allow a
compartment-specific handler to catch and handle the violation signal, ensuring the system can
continue to operate---thus, meeting the improved availability design goal.
We have reproduced CVE-2018-16526, and developed an exploit for it as a
proof-of-concept example for vulnerability mitigation and improved availability evaluation.
This CVE is very severe, and similar to other CVEs (as far as reproduction and exploitation);
it is also a base on which others can build.

\subsection{Case Study: Safety-Critical Automotive ECU}
\label{usecase:automotive}

\begin{figure}[]
    \centering
    \includegraphics[width=\columnwidth,height=4cm]{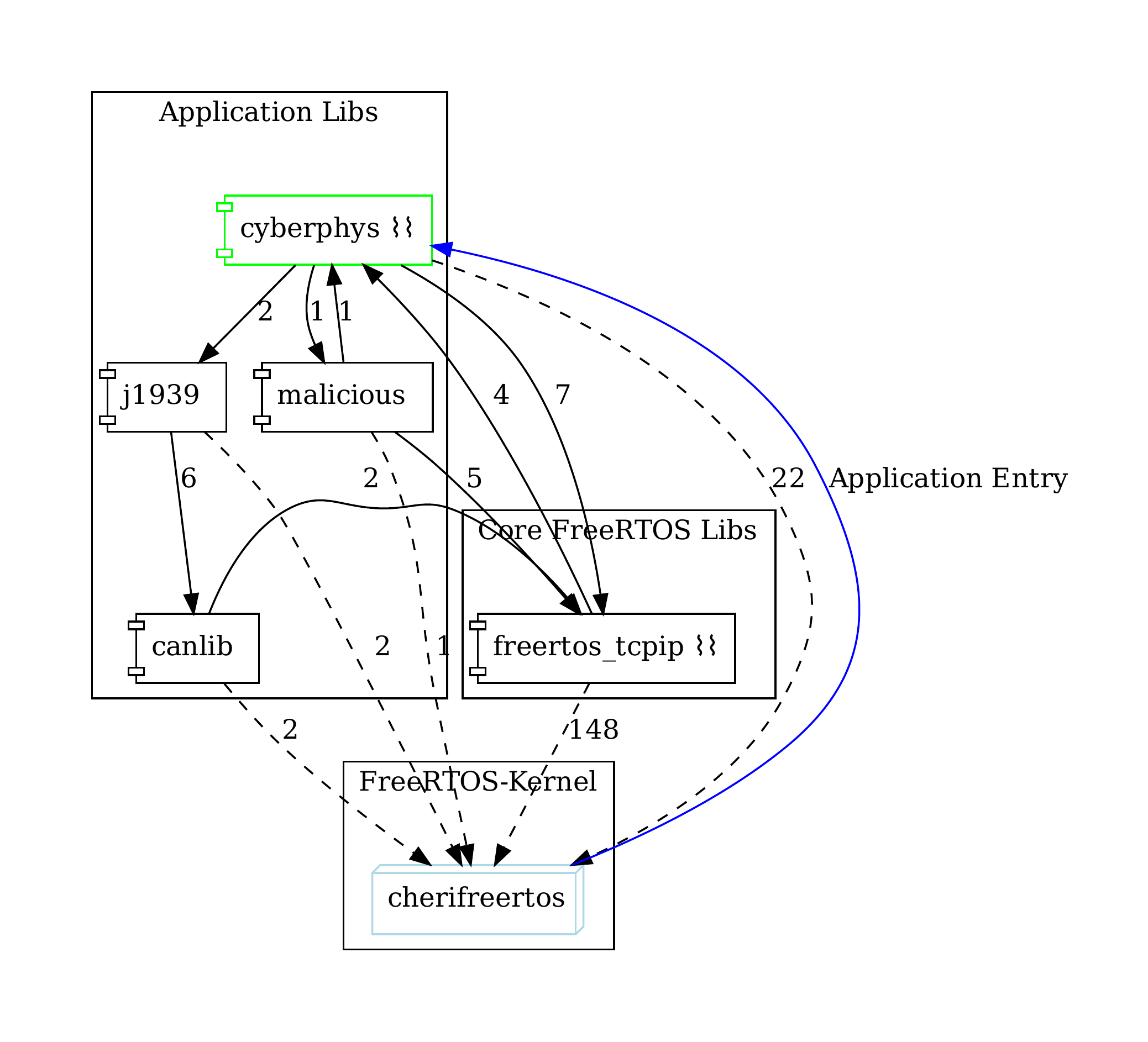}
    \caption{Automotive ECU compartmentalized graph.}
    \label{fig:cyberphys_generated}
\end{figure}

The Automotive ECU demonstrator is an existing FreeRTOS-based system (not created by us) on which
we apply the CompartOS model to evaluate availability (partial recovery) and compatibility (no source-code changes) of a real-time safety-critical system. The demonstrator (see Figure~\ref{fig:cyberphys_generated}) imitates a modern car,
where multiple microcontrollers are responsible for
various tasks and communicate with each other over a CAN network~\cite{canprotocol2021}. This ECU
runs FreeRTOS and contains a buffer-overflow vulnerability in the SAE J1939~\cite{j19392021}
protocol stack. An attacker could make use of that and carefully craft a J1939 packet that leads
to a buffer overflow and remote code execution (e.g., disabling reading the brake pedal
position).




\subsubsection{Fault Handling}
\label{sec:eval_fault}

The ECU demo is a good use case for evaluating a few fault handling techniques within CompartOS as it is
a safety-critical system that is also subject to real-time
requirements. It does integrate both critical and less critical compartments with different
attributes and requirements.
We evaluate fault handling in the three following compartments and report the lesson and conclusions learned
for each scenario.

%

\subsubsection{J1939}
The J1939 compartment is safety-critical and needs to be reliable and running all the time. It is
used to send brakes and steering commands with real-time requirements; the main loop delay is
\textbf{\textit{50 milliseconds}}; thus, commands should be detected and handled in less than that.

\textbf{Return-Error:}
We experimented with a per-compartment fault handler that simply drops the malicious command and
returns to the caller with an error. The entire fault handling and return take
approximately \textbf{\textit{30 microseconds}}, which will not have any effect on the car's
steering
logic; therefore, the real-time guarantees are
maintained. On the other hand, a non-compartmentalized system may be forced to perform a full
restart as a brute-force recovery mechanism, which takes at least 2 seconds (the boot time of
FreeRTOS and the Automotive ECU application)---an unacceptable latency for a safety-critical car
brake system. However, Return-Error handling could cause memory exhaustion and \gls{DoS} if the attack is
recurrent as some buffers are allocated but not freed in that case.

\textbf{Custom-Handler:}
To address the issue of \gls{DoS} incurred in Return-Error, we implemented a Custom-Handler, which
requires knowledge of the design and implementation of the compartment itself, and assumes a
faulting compartment would return an error code. It then takes further actions to free the buffer
itself, thus not causing \gls{DoS} or memory exhaustion. This is a more robust and secure strategy, but
it requires a good knowledge of the underlying compartments, potential faults (to TRY-CATCH), and
adds some implementation efforts. The custom handler for J1939 took 10 LoC changes that are mostly
a new function to free the allocated canlib's buffer. The overall fault handing took
\textbf{\textit{60 microseconds}}, which includes an addition of 30 microseconds to free the buffer, which is
completely acceptable and meets the real-time requirements.

\textbf{Micro-reboot:}
The J1939 compartment has some attributes that made it possible to be micro-rebooted while meeting
real-time requirements. The compartment is stateless; it only performs services, allocates buffers,
processes and decodes the J1939 packets and finally returns.
Further, it is small and simple enough to perform a snapshot of its ELF sections and roll them back to
a known start state on faults, effectively performing a micro-reboot. The entire compartment, including
the captable, .text, .data, etc., fit in less than 4 KiB.
Since .text and .rodata cannot be written (due to CHERI permissions), only .bss and .data need to be
rolled back to a start state. The process of doing so takes \textbf{\textit{83 microseconds}}.
No code changes were required at all. Rolling back all the ELF sections (though unnecessary) takes 220 microseconds.

\subsubsection{Malicious}
The \textit{Malicious} compartment represents a non-critical compartment that might have been dynamically loaded
or taken control of. This threat model is not addressed in most secure embedded systems like TockOS
and ACES. This compartment could be something like a web browser, FTP server, or music player; all
are non-critical to the car steering system but may have zero-day vulnerabilities that might be
exploited. Malicious could try to attack confidentiality and integrity, and that will be prevented
by virtue of capability-based security offered by CHERI/CompartOS. More importantly, Malicious could try to perform \gls{DoS} attacks
by faulting frequently or performing recursive calls, thus preventing critical compartments from
executing or affecting their real-time requirements.
A few fault handling techniques could be applied here as Malicious is non-critical and does
not have critical compartments relying on it. However, in such a scenario, the compartment could
be simply killed or suspended for good until an appropriate action is performed (e.g., software updates).

\textbf{Compartment-Kill}
Malicious has faulting instruction and can be killed. Killing a compartment is as
straightforward as invalidating its root capability table register (CGP).
Malicious tries to create sockets and send data over a port. The first caught violation attempt
takes 14.4 milliseconds which may affect the real-time requirements of the main loop as it is part
of its path. Once caught by CHERI, the entire fault handling
technique (by CompartOS) takes \textbf{\textit{22 microseconds}} which meets real-time requirements. Further
attempts to invoke the compartment immediately fault as CGP is invalid, and that takes 22 microseconds or
less. This prevents the Malicious compartment from accessing any external or internal functions or
data capabilities and protects against \gls{DoS} attempts. Finally, this technique did not require any
source-code changes at all on the application level, thus maintaining compatibility.

\subsubsection{TCP/IP}
The TCP/IP stack is the most complex compartment in this demo and likely most other RTOSes as well.
It does have many resources, dynamically allocates memory, integrates network device drivers, and
has many dependencies on it. Thus, a single vulnerability in the TCP/IP stack can massively affect
the availability and overall security of the system. We have reproduced CVE-2018-16526, discussed
earlier, and demonstrated a real attack that exploits it to understand the implications associated
with fault handling in such scenarios.
First, in a non-secure system, this vulnerability could cause integrity and confidentially
violations (e.g., remote code execution), which might take control of steering the car or crash it.
Second, in a PURECAP system that is non-compartmentalized, the buffer overflow will be detected by
CHERI, but this will trigger an exception and put the entire system on halt, thus affecting its
availability and causing a crash.
CompartOS is uniquely able to identify the faulting compartment, its boundaries, and dependencies
which allows it to take further fault handling actions.
Only a custom fault handler could work for the TCP/IP stack as opposed
to other fault handling techniques discussed before.
Thus, we chose
to experiment with a custom fault handler that restarts the TCP/IP stack. The following
implementation actions were required:
1) reset some of the TCP/IP stack state,
2) free memory resources and kill threads created by the compartment,
3) notify the dependent compartments that the TCP/IP stack is going to be restarted,
4) dependent compartments need to stop using the sockets and  the TCP/IP compartments,
5) reinitialize the TCP/IP stack,
6) dependent compartments need to recreate sockets and can resume using the TCP/IP compartment.
In our evaluation, it took 1 second to
perform a complete restart (in the background), while other critical components such as
\textit{cyberphys} kept functioning and meeting their real-time requirements. We compare that
against non-compartmentalized \textbf{INSECURE-STATIC} and \textbf{PURECAP} systems that take at least 2
seconds to perform a complete restart of the whole system that stops all safety-critical
compartments.

\section{Conclusions and future work}
We have described CompartOS as a lightweight linkage-based compartmentalization model for existing, large,
complex, mainstream embedded systems. Through our CheriFreeRTOS implementation of
CompartOS, we evaluated performance, compatibility, availability, scalablity, and security of both
micro- and macro- benchmarks. We showed that CompartOS outperforms all MPU-based security models
and could protect against (future) exploits that MPUs cannot detect. Further, CompartOS maintains
source-code compatibility while compartmentalizing a large code-base of hundreds or millions LoC
with hundreds of compartments and resources, unlike MPU-based solutions. Still, CompartOS
also preserves the traditional determinism requirement of RTOSes, unlike MMUs.
Finally, we showed how availability of a safety-critical system can be improved
through a few fault handling mechanisms and partial recovery, while meeting real-time requirements.
Our future work will focus on the formalization and evaluation of security
policies across specific security applications and standards such as ARINC 653~\cite{arinc22wiki}.
The CompartOS linkage-based model could also be applied to other (embedded) operating systems,
boot loaders, and even general purpose operating systems. We will also explore
how CompartOS can be applied to compartmentalize Linux Kernel Modules.

\section{Acknowledgments}
This work was supported by the Defense Advanced Research
Projects Agency (DARPA) under contracts HR0011-18-C-0016 (``ECATS'')
and HR0011-18-C-0013 (``BESSPIN''). The
views, opinions, and/or findings contained in this report are
those of the authors and should not be interpreted as representing the official views or policies of the Department of
Defense or the U.S. Government. We also acknowledge the
Gates Cambridge Trust, EPSRC grant EP/R513180/1, Arm
Limited, and Google. Approved for Public Release, Distribution Unlimited.

\bibliographystyle{plain}
\bibliography{compartos}

\end{document}